\documentclass[twocolumn,epjc3]{svjour3}
\smartqed 
\RequirePackage{graphicx} 
\RequirePackage[numbers,sort&compress]{natbib}
\journalname{Eur. Phys. J. C}
\usepackage{amsmath,latexsym,epsfig,rotate,amssymb}

\begin{document}
\title{Contribution of vector resonances to the  ${\bar B}_d^0\to {\bar K}^{*0}\,\mu^+\,\mu^-$ decay }
\author{Alexander Yu. Korchin\thanksref{e1,addr1,addr2}        \and
        Vladimir A. Kovalchuk\thanksref{e2,addr1}}
\thankstext{e1}{e-mail: korchin@kipt.kharkov.ua}
\thankstext{e2}{e-mail: koval@kipt.kharkov.ua}
\institute{NSC `Kharkov Institute of Physics and Technology',
61108 Kharkiv, Ukraine\label{addr1}          \and
          V.N.~Karazin Kharkiv National University, 61022 Kharkiv, Ukraine\label{addr2}
       }
       \date{Received: date / Accepted: date}
\maketitle

\begin{abstract}
The fully differential angular distribution for the rare flavor-changing
neutral current decay ${\bar B}_d^0 \to {\bar K}^{*0} \, (\to K^{-} \pi^+)\,
\mu^+\,\mu^- $ is studied. The emphasis is placed on accurate treatment of the
contribution from the processes $\bar{B}_d^0 \to \bar{K}^{*0} (\to K^{-}\,
\pi^+) \, V $ with intermediate vector resonances $V$ = $\rho(770)$, $
\omega(782)$, $\phi(1020)$, $J/ \psi$, $\psi(2S)$, $\ldots$ decaying into the
$\mu^+ \mu^-$ pair.
The dilepton invariant-mass dependence of the branching ratio, longitudinal
polarization fraction $f_L$ of the ${\bar K}^{*0}$ meson, and forward-backward
asymmetry $A_{\rm FB}$ is calculated and compared with data from Belle, CDF and
LHCb. It is shown that inclusion of the resonance contribution may considerably
modify the branching ratio, calculated in the SM without resonances, even in
the invariant-mass region far from the so-called charmonia cuts applied in the
experimental analyses. This conclusion crucially depends on values of the
unknown phases of the $B^0 \to K^{*0} \, J/\psi$ and $B^0 \to K^{*0} \,
\psi(2S)$ decay amplitudes with zero helicity.

\end{abstract}

\section{\label{sec:Introduction}Introduction}

The investigation of the rare decay \[{\bar B}_d^0 \to {\bar
K}^{*0}\,(\to K^{-}\,\pi^+)\,\mu^+ \,\mu^- \] induced by the
flavor-changing neutral current (FCNC) transition $b\to s
\mu^+\mu^-$ is an important test of the standard model (SM) and
its extensions (see \cite{Antonelli:2009} for a review). The
phenomenology of this decay mode has been discussed by many
authors, e.g. see Refs.~\cite{Melikhov:1998, Kruger:2000,
Ali:2000, Kim:2000, Ali:2002, Kruger:2005, Lunghi:2007,
Bobeth:2008, Bobeth:2010, Bobeth:2011, Egede:2008,
Altmannshofer:2009, Korchin:2010, Egede:2010, Lunghi:2010,
Bharucha:2010, Alok:2010, Alok:2011, Alok:2011a,
DescotesGenon:2011, Becirevic:2012, Altmannshofer:2011,
Korchin:2011, Matias:2012, Das:2012, Das:2012a}.

This decay takes place in a very wide region of dimuon invariant
mass squared, $q^2=(q_++q_-)^2$, namely $4m_\mu^2\leq q^2\leq
q^2_{\rm max}=(m_B-m_{K^{*}})^2$. The light vector resonances
$\rho(770)$, $\omega(782)$, $\phi(1020)$ (and their radial
excitations) and the $c \bar{c}$ resonances $ J/\psi$, $\psi(2S)$
(and higher states) are also located in this region.
Thus at $q^2 \approx m_V^2$ the decay ${\bar B}_d^0 \to {\bar K}^{*0} \,
\mu^+\, \mu^- $ can go through the hadronic weak decay ${\bar B}_d^0 \to {\bar
K}^{*0} \, V$, followed by the dimuonic annihilation of vector meson $V$. All
resonances with $m_V<m_B-m_{K^{*}}$ make a contribution to this mechanism.
Therefore both the nonresonant and resonant parts can contribute to the total
amplitude of the decay ${\bar B}_d^0 \to {\bar K}^{*0} \, \mu^+\, \mu^- $.

Main attention in literature has been paid to description of the nonresonant
amplitude of the decay ${\bar B}_d^0 \to {\bar K}^{*0} \, \mu^+\, \mu^- $ in
the region $1\, {\rm GeV^2} \leq q^2\leq 6\, {\rm GeV^2}$. In this region,
using the QCD factorization (QCDf) \cite{Beneke:1999, Beneke:2000}, one can
perform a systematical calculation of non-factorizable corrections  to ``naive
factorization approximation'' (NFA) and spectator effects \cite{Beneke:2001a,
Beneke:2005}. At larger dimuon masses, at about $q^2 > 14\,{\rm GeV^2}$, the
QCDf and the light-cone sum rules (LCSR) methods are not applicable. For the
estimation of non-factorizable corrections, an operator product expansion in
powers of $1 / \sqrt {q^2}$ can be used \cite{Grinstein:2004, Beylich:2011}. In
the region $10$ GeV$^2$ $\lesssim q^2 \lesssim 13 $ GeV$^2$ the
non-factorizable effects due to soft-gluon emission have been included
in~\cite{Khodjamirian:2010}.

Often the resonant contribution to amplitudes of rare decays of
$B$-meson is modeled in terms of the Breit-Wigner functions for
the
resonances~\cite{Deshpande:1989,Lim:1989,Ali:1991,Ligeti:1996,Ligeti:1998}.
In these references the resonance corrections are added to the
perturbative loops of charm quarks.
Note an original approach of Ref.~\cite{Kruger:1996} for the inclusive $B \to
X_s \, \ell^+ \ell^-$ process, in which dispersion relation exploiting
experimentally measured cross section $\sigma(e^+ e^- \to {\rm hadrons})$ has
been applied to account for the resonance terms (see also~\cite{Beneke:2009}).

These approaches more often apply to the inclusive decays $B \to X_s \, \ell^+
\ell^-$ and use information on the $B \to X_s \, J/\psi$ and $B \to X_s \,
\psi(2S)$ branching ratios for description of the resonant contribution.
Sometimes, such approaches are extended to the exclusive decays $B \to K (K^*)
\, \ell^+ \ell^-$ \cite{Ali:2000, Ligeti:1996,Ligeti:1998}, in which the
branching ratios for exclusive decays $B \to K (K^*) \, J/\psi$ and $B \to K
(K^*) \, \psi(2S)$ are used. In these studies, carried out in framework of the
NFA, additional factors $k_V$ are introduced into the resonant terms to adjust
the branching ratios for the decays, for instance,
\begin{eqnarray*}
BR \, (B \to K^* \, V  &\to& K^* \, \ell^+ \ell^-) \\ & = & BR\, (B \to K^* \,
V) \, BR \, ( V \to \ell^+ \ell^-)\,,
\end{eqnarray*}
where the right-hand side is taken from experiment.

Recall that, in general, the process $B \to K^* \, V $ is characterized not
only by the branching ratio. The decay of a $B_d^0$ meson into a pair of vector
mesons, $ B_d^0 \to K^{*0} \, V$, is described by three complex amplitudes
\cite{Valencia:1989}. In the transverse basis \cite{Dunietz:1991, Dighe:1996},
these decay amplitudes correspond to linearly polarized states of vector
mesons, which are polarized either longitudinally ($0$) or transversely to the
direction of their motion, being polarized in parallel ($\|$) or perpendicular
($\perp$) to each another.
Overall, six real parameters describe three complex amplitudes
$A_0^V$, $A_\|^V$, and $A_\perp^V$. They could be chosen to be,
for example, the branching ratio, $|A_0^V|^2$, $|A_\|^V|^2$, ${\rm
arg}(A_\|^V/A_0^V)$, ${\rm arg}(A_\perp^V/A_0^V)$ and one overall
phase ${\rm arg}(A_0^V)$. The phase convention is arbitrary for an
isolated decay $ B_d^0 \to K^{*0} \, V$.
Sometimes, this phase is chosen zero, ${\rm arg}(A_0^V)=0$.
However, for certain $B$ decays, this phase can produce meaningful
and observable effects, such as for $B\to V\,K^*_J$ with
$J=0,1,2,\ldots$.

For example, in the decay  $ B_d^0 \to K^{*0} \, \phi$, the phase of the
amplitude $A_0^\phi$ has been measured with respect to the phase of the
amplitude $A_{00}^\phi$ of the decay $ B_d^0 \to K_0^*(1430)^0 \, \phi$ and is
equal to ${\rm arg}(A_{00}^\phi/A_0^\phi)=2.82\pm0.15\pm0.09$ \cite{Babar:2007,
Babar:2008}. For the other vector resonances $V$, the corresponding relative
phase has not been measured so far.

At present in decay modes to the light resonances, $ B_d^0 \to K^{*0} \,
\rho(770)$ and $ B_d^0 \to K^{*0}$ $\omega(782)$, only the branching ratio and
longitudinal polarization fraction of the $K^*$ meson are measured, while the
decays to radial excitations, $\omega(1420)$, $\rho(1450)$, $ \omega(1650)$,
$\rho(1700)$, $\cdots $, have not been observed. At the same time,  all
amplitudes of the decay $ B_d^0 \to K^{*0} \, \phi(1020)$ are known from
experiment, while there is no information on $B$ decays to radial excitations
of $\phi(1020)$.
For the $ B_d^0 \to K^{*0} \, J/\psi$ and $ B_d^0 \to K^{*0} \,
\psi(2S)$ decays, the full angular analysis has been performed. As
for the decays to the higher states, $\psi(3770)$, $\psi(4040)$,
$\psi(4160)$, $\psi(4415)$, the experimental information is
absent.

In the present paper for description of resonant contribution to the four-body
decay ${\bar B}_d^0 \to K^{-}\, \pi^+ \, \mu^+ \,\mu^- $ the available
information on the helicity amplitudes for $ B \to K^* \, V$ is used.
The fully differential angular distribution over the three angles
and dimuon invariant mass is analyzed in the whole region $ 4
m_\mu^2 \le q^2 \le (m_B-m_{K^*})^2$. The amplitude of this decay
consists of the nonresonant amplitude in the SM model and the
resonant amplitude. For the first amplitude we use the NFA, in
which hadronic matrix elements are parameterized in terms of form
factors \cite{Ball:2005}, and the Wilson coefficients are taken in
the next-to-next-to-leading order (NNLO) approximation.

The resonant amplitude is expressed in terms of the invariant
amplitudes $S_{1,2,3}^V$ for the decays $ B_d^0 \to K^{*0} \, V$.
The information on the latter is taken from experiment if
available, or from theoretical estimations.
As mentioned above, the  phase $\delta_0^V \equiv {\rm arg}(A_0^V)$ for an
isolated decay $ B_d^0 \to K^{*0} \, V$ is arbitrary. This phase may produce
observable effects in the decay ${\bar B}_d^0 \to \bar{K}^{*0} \, \mu^+ \,\mu^-
$ via the interference with the nonresonant amplitude. We investigate influence
of the phases $\delta_0^V$ for each resonance $V$ on the differential branching
ratio, longitudinal polarization fraction $f_L$ of $K^*$ and forward-backward
asymmetry $A_{\rm FB}$.

We also study two aspects of the resonant amplitude. The first one is related
to the fact that the vector mesons are off their mass shells, therefore an
off-mass-shell extension of the on-mass-shell $ B_d^0 \to K^{*0} \, V$
amplitudes is proposed. The second one is the choice of the
vector-meson-dominance (VMD) model which describes the transition vertex $V \,
\gamma $. We use two versions of the VMD model (called further VMD1 and VMD2)
which result in rather different $V \, \gamma $ vertices. In particular, in the
VMD2 model, the $V \, \gamma $ vertex is suppressed compared to the VMD1 vertex
in the region $q^2 \ll m_V^2$.

Results of the present calculations are compared with the recent
data from Belle, CDF and LHCb experiments. Usually in these
experiments the $c \bar{c}$ resonance contributions are removed by
putting cuts on the invariant dimuon mass near the resonance mass
$q^2=m^2_V$. This assumption is used in the analyses of all
ongoing and planned experiments.

The paper is organized as follows. In Section~\ref{subsec:angle
distribution} the fully differential angular distribution is
discussed. Nonresonant and resonant amplitudes in the transverse
basis are specified in Section~\ref{subsec:resonances}. Results
for the dependence of observables on the invariant mass squared
are presented in Section~\ref{sec:results}. Conclusions are drawn
in Section~\ref{sec:conclusions}. In~\ref{subsec:vector mesons}
calculation of the $B^0 \to K^{*0} V$ amplitudes for the
off-mass-shell vector meson $V$ is described.

\section{\label{sec:formalism} Angular distributions and
amplitudes for the ${\bar B}_d^0\to {\bar K}^{*0}\,\mu^+\,\mu^-$
decay }

\subsection{ \label{subsec:angle distribution}
Differential decay rate}

\begin{figure}
\centerline{\includegraphics[width=.45\textwidth]{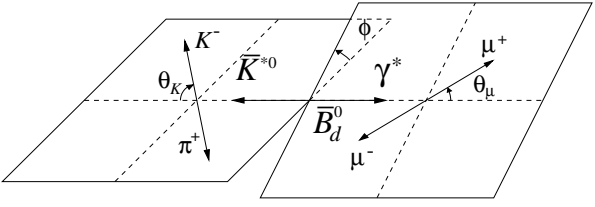}}
\caption{Definition of helicity angles $\theta_\mu$, $\theta_K$,
and $\phi$, for the decay ${\bar B}_d^0\to {\bar
K}^{*0}\,\mu^+\,\mu^-$.} \label{fig1}
\end{figure}

The decay ${\bar B}_d^0\to {\bar K}^{*0}\,\mu^+\,\mu^-$, with
${\bar K}^{*0} \to K^- \pi^+$ on the mass shell~\footnote{This
means the narrow-width approximation for the ${\bar K}^{*0}$
propagator: \ $(k^2 - m_{K^*}^2 + im_{K^*} \Gamma_{K^*})^{-1}
\approx -i \pi \delta(k^2 - m_{K^*}^2) $.}, is completely
described by four independent kinematic variables: the dimuon
invariant mass squared, $q^2$, and the three angles $\theta_\mu$,
$\theta_K$, $\phi$. In the helicity frame (Fig.~\ref{fig1}), the
angle $\theta_\mu\,(\theta_K)$ is defined as the angle between the
directions of motion of $\mu^+\,(K^-)$ in the $\gamma^*\,({\bar
K}^{*0})$ rest frame and the $\gamma^*\,({\bar K}^{*0})$ in the
${\bar B}_d^0$ rest frame. The azimuthal angle $\phi$ is defined
as the angle between the decay planes of $\gamma^*\to
\mu^+\,\mu^-$ and ${\bar K}^{*0} \to K^- \pi^+$ in the ${\bar
B}_d^0$ rest frame. The differential decay rate in these
coordinates is given by
\begin{eqnarray}
&&\frac{d^4\,\Gamma}{d\hat{q}^2 d\cos\theta_{\mu} d\cos\theta_K
d\phi}\nonumber
\\ &&=\beta_{\mu}\,m_B\,N^2\hat{q}^2\sqrt{\hat{\lambda}}
\frac{9}{64\,\pi}\sum_{k=1}^{11}a_{k}(q^2)g_{k}(\theta_{\mu},
\theta_K,\phi)\,,\label{eq:001}
\end{eqnarray}
where the angular terms $g_k$ are defined as
\begin{eqnarray}
&&g_1=4\sin^2\theta_{\mu}\cos^2\theta_K\,,g_2=(1+\cos^2\theta_{\mu})\sin^2\theta_K\,,\nonumber
\\ &&g_3=\sin^2\theta_{\mu}\cos2\phi\sin^2\theta_K\,,\nonumber
\\
&&g_4=-2\sin^2\theta_{\mu}\sin^2\theta_K\sin2\,\phi\,,\nonumber\\
&&g_5=-\sqrt{2}\sin2\,\theta_{\mu}\sin2\,\theta_K\cos\phi\,,\nonumber
\\ && g_6=-\sqrt{2}\sin2\,\theta_{\mu}\sin2\,\theta_K\sin\phi\,,
g_7=4\cos\theta_{\mu}\sin^2\theta_K\,,\nonumber \\
&&g_8=-2\sqrt{2}\sin\theta_{\mu}\sin2\,\theta_K\cos\phi\,,\nonumber
\\
&&g_9=-2\sqrt{2}\sin\theta_{\mu}\sin2\,\theta_K\sin\phi\,,\nonumber
\\ && g_{10}=2\cos^2\theta_K\,,g_{11}=\sin^2\theta_K\,,\nonumber
\end{eqnarray}
and the amplitude terms $a_k$ as
\begin{eqnarray}
&&a_1=\beta_{\mu}^2\,|A_{0}|^2\,,\:
a_2=\beta_{\mu}^2\,(|A_{\|}|^2+|A_{\perp }|^2)\,,\nonumber \\
&&a_3=\beta_{\mu}^2\,(|A_{\perp }|^2-|A_{\|}|^2)\,,\:
a_4=\beta_{\mu}^2\,{\rm Im}(A_{\|}A_{\perp
}^*)\,, \nonumber \\
&&a_5=\beta_{\mu}^2\,{\rm Re}(A_{0}A_{\|}^*)\,,\:
a_6=\beta_{\mu}^2\,{\rm Im}(A_{0}A_{\perp
}^*)\,,\nonumber \\
&&a_7=\beta_{\mu}\,{\rm Re}(A_{\|L}A_{\perp L}^*-A_{\|R}A_{\perp
R}^*)\,,\nonumber \\ &&a_8=\beta_{\mu}\,{\rm Re}(A_{0L}A_{\perp
L}^*-A_{0R}A_{\perp R}^*)\,,\nonumber \\
&&a_9=\beta_{\mu}\,{\rm Im}(A_{0L}A_{\| L}^*-A_{0R}A_{\|
R}^*)\,,\nonumber \\
&&a_{10}=(1-\beta_{\mu}^2)(|A_{0L}+A_{0R}|^2+|A_t|^2)\,,\nonumber
\\&&a_{11}=(1-\beta_{\mu}^2)(|A_{\|L}+A_{\|R}|^2+|A_{\perp
L}+A_{\perp R}|^2)\,,\nonumber
\end{eqnarray}
where $\beta_{\mu}\equiv\sqrt{1-4m_{\mu}^2/q^2}$, $m_{\mu}$ is the
mass of the muon, $m_B$ is the mass of the $B^0_d$ meson,
$\hat{q}^2\equiv q^2/m_B^2$, and
\begin{equation}
A_i A^*_j \equiv A_{i L}(q^2) A^*_{jL}(q^2)+ A_{iR}(q^2)
A^*_{jR}(q^2) \nonumber \, .
\end{equation}
Here $i,j  = (0, \|, \perp )$, the $a_k$ dependent on products of
the six transversity amplitudes $A_{0L(R)}$, $A_{\|L(R)}$ and
$A_{\perp L(R)}$, where $L$ and $R$ refer to the chirality of the
leptonic current, as well as the seventh transversity amplitude
$A_t$. The latter amplitude is related to the time-like component
of the virtual gauge boson, which does not contribute to the decay
rate in the case of massless leptons and can be neglected if the
lepton mass is small in comparison to the invariant-mass of the
leptonic pair. Further, $\hat{\lambda}\equiv\lambda(1,\hat{q}^2,
\hat{m}_{K^*}^2)=(1-\hat{q}^2)^2-2(1+\hat{q}^2)\hat{m}_{K^*}^2+\hat{m}_{K^*}^4$,\
$\hat{m}_{K^*}\equiv m_{K^*}/m_B$, where $m_{K^*}$ is the mass of
the $K^{*0}$ meson, and
\[N=|V_{tb}V_{ts}^*|\frac{G_F m_B^2 \alpha_{\rm em}}{32 \,\pi^2 \sqrt{3\,
\pi}}\,.\]
Here, $V_{ij}$ are the Cabibbo-Kobayashi-Maskawa (CKM) matrix
elements \cite{Cabibbo:1963, Kobayashi:1973}, $G_F$ is the Fermi
coupling constant, $\alpha_{\rm em}$ is the electromagnetic
fine-structure constant.

The longitudinal, parallel, and perpendicular partial widths are
given, respectively, by
\begin{eqnarray}
\frac{d\,\Gamma_L}{d\hat{q}^2}&=&m_B N^2\beta_\mu
\hat{q}^2\sqrt{\hat{\lambda}}\Bigl(\beta_\mu^2|A_0|^2\nonumber \\
&&+\frac{3m_\mu^2}{q^2}\bigl(|A_t|^2
+|A_{0L}+A_{0R}|^2\bigr)\Bigr)\,.\label{eq:002}
\end{eqnarray}
\begin{eqnarray}
\frac{d\,\Gamma_\|}{d\hat{q}^2}&=&m_B N^2\beta_\mu
\hat{q}^2\sqrt{\hat{\lambda}}\Bigl(\beta_\mu^2|A_{\|}|^2\nonumber \\
&&+\frac{3m_\mu^2}{q^2}|A_{\|L}+A_{\|R}|^2\Bigr)\,.\label{eq:003}
\end{eqnarray}
\begin{eqnarray}
\frac{d\,\Gamma_\perp}{d\hat{q}^2}&=&m_B N^2\beta_\mu
\hat{q}^2\sqrt{\hat{\lambda}}\Bigl(\beta_\mu^2|A_{\perp}|^2\nonumber \\
&&+\frac{3m_\mu^2}{q^2}|A_{\perp L}+A_{\perp
R}|^2\Bigr)\,.\label{eq:004}
\end{eqnarray}
The familiar muon-pair invariant-mass spectrum for ${\bar
B}_d^0\to {\bar K}^{*0}\,\mu^+\,\mu^-$ decay can be recovered
after integration over all angles as
\begin{eqnarray}
\frac{d\,\Gamma}{d\hat{q}^2}&=&\frac{d\,\Gamma_L}{d\hat{q}^2}+\frac{d\,\Gamma_\|}{d\hat{q}^2}+\frac{d\,\Gamma_\perp}{d\hat{q}^2}\,.
\label{eq:005}
\end{eqnarray}
The fraction of $K^*$ meson polarization is \ $[i=(L, \|, \perp)]$
\[f_i(q^2)=\frac{d\,\Gamma_i}{d\hat{q}^2}/\frac{d\,\Gamma}{d\hat{q}^2}\,.
 \]
Integrating Eq.~(\ref{eq:001}) over the variables $\cos\theta_\mu$
and $\phi$, we obtain
\begin{eqnarray}
\frac{d^2\,\Gamma}{d\hat{q}^2d\cos\theta_K}&=&\frac{d\,\Gamma}{d\hat{q}^2}\,\Bigl(\frac{3}{2}f_L
\cos^2\theta_K \nonumber \\
&&+\frac{3}{4} (1-f_L)(1-\cos^2\theta_K)\Bigr)\,.
 \label{eq:006}
\end{eqnarray}
Integration of Eq.~(\ref{eq:001}) over $\cos\theta_K$ and $\phi$
yields
\begin{eqnarray}
&&\frac{d^2\,\Gamma}{d\hat{q}^2d\cos\theta_\mu}=m_B N^2\beta_\mu
\hat{q}^2\sqrt{\hat{\lambda}}\,\frac{3}{8}\Bigl(2\,\beta_\mu^2|A_0|^2\sin^2\theta_\mu \nonumber \\
&&+\beta_\mu^2\left(|A_{\|}|^2+|A_{\perp}|^2\right)\left(1+\cos^2\theta_\mu
\right)+\frac{4m_\mu^2}{q^2}\bigl(|A_t|^2 \nonumber \\
&&+|A_{0L}+A_{0R}|^2 +|A_{\|L}+A_{\|R}|^2\nonumber \\
&&+|A_{\perp L}+A_{\perp R}|^2\bigr)\Bigr)+\frac{d{A}_{\rm FB}^{
(\mu)}}{d\hat{q}^2}\cos\theta_\mu \,,
 \label{eq:007}
\end{eqnarray}
where $d{A}_{\rm FB}^{\rm (\mu)}/d\hat{q}^2$ is the muon
forward-backward asymmetry,
\begin{eqnarray}
&&\frac{d{A}_{\rm FB}^{
(\mu)}}{d\hat{q}^2}\equiv\int\limits_{-1}^{1}{\rm
sgn}(\cos\theta_\mu)\frac{d^2\,\Gamma}{d\hat{q}^2d\cos\theta_\mu}\:d\cos\theta_\mu\nonumber \\
&=&m_B N^2\beta^2_\mu
\hat{q}^2\sqrt{\hat{\lambda}}\,\frac{3}{2}{\rm
Re}\left(A_{\parallel\,L}\, A_{\perp\,L}^*-A_{\parallel\,R}\,
A_{\perp\,R}^*\right)\, , \label{eq:008}
\end{eqnarray}
and the normalized forward-backward asymmetry \\$d{\bar A}_{\rm
FB}^{(\mu)}/d\hat{q}^2$ is given as
\begin{equation} \frac{d{\bar A}_{\rm FB}^{
(\mu)}}{d\hat{q}^2}\equiv\frac{d A_{\rm FB}^{
(\mu)}}{d\hat{q}^2}/\frac{d \Gamma}{d\hat{q}^2}\,. \label{eq:009}
\end{equation}
Finally, the one-dimensional angular distribution in the angle
$\phi$ between the lepton and meson planes takes the form
\begin{eqnarray} \label{eq:010}
\frac{d^2\,\Gamma}{d\hat{q}^2d\phi}&=&\frac{1}{2\pi}\frac{d\,\Gamma}{d\hat{q}^2}
\Bigl(1+\frac{1}{2}\bigl(1-f_L\bigr)A^{(2)}_{\rm
T}\cos2\phi\nonumber \\&& -A_{\rm Im}\sin2\phi\Bigr) ,
\end{eqnarray}
\begin{equation} \label{eq:011}
A^{(2)}_{\rm T}\equiv
\left(\frac{d\,\widetilde{\Gamma}_\perp}{d\hat{q}^2}-\frac{d\,\widetilde{\Gamma}_\|}{d\hat{q}^2}\right)/
\left(\frac{d\,\Gamma_\perp}{d\hat{q}^2}+\frac{d\,\Gamma_\|}{d\hat{q}^2}\right)\,,
\end{equation}
\begin{equation} \label{eq:012}
\frac{d\,\widetilde{\Gamma}_{\perp (\|)}}{d\hat{q}^2}=m_B
N^2\beta^3_\mu \hat{q}^2\sqrt{\hat{\lambda}}\,|A_{\perp (\|)}|^2
\,,
\end{equation}
\begin{equation} \label{eq:013}
A_{\rm Im}\equiv m_B N^2\beta^3_\mu
\hat{q}^2\sqrt{\hat{\lambda}\,}{\rm
Im}(A_\|A^*_\perp)/\frac{d\,\Gamma}{d\hat{q}^2} ,
\end{equation}
where the asymmetry $A^{(2)}_{\rm T}(q^2)$ is sensitive to new
physics from right-handed currents beyond the standard model, and
the amplitude $A_{\rm Im}(q^2)$ is sensitive to complex phases in
the hadronic matrix elements. Sometimes $A^{(2)}_{\rm T}(q^2)$ is
called transverse asymmetry~\cite{Kruger:2005}.


\subsection{\label{subsec:resonances}
 Resonant and nonresonant transverse amplitudes }

\begin{figure}[tbh]
\centerline{\includegraphics[width=.45\textwidth]{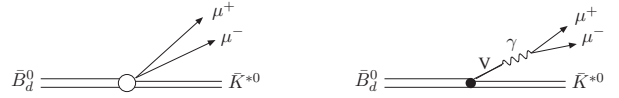}}
\caption{Nonresonant and resonant contributions to the decay
amplitude. } \label{fig:resonances}
\end{figure}

The effects of the long-distance contribution from the decays
${\bar B}_d^0\to {\bar K}^{*0}\,V$, where $V=\rho^0$, $\omega$,
$\phi$, $J/\psi(1S)$, $\psi(2S)$, $\ldots$ mesons, followed by
$V\to \mu^+\,\mu^-$ in the decay ${\bar B}_d^0\to {\bar
K}^{*0}\,\mu^+\,\mu^-$ are included through the VMD approach, as
shown in Fig.~\ref{fig:resonances}.
There is no unique way of introducing the $V \, \gamma$ transition, and one can
use various versions of VMD models which yield different $V \, \gamma$
transition vertices. In one of VMD models (see \cite{Feynman}, chapter~6)
\begin{equation} \langle \gamma (q); \, \mu | V (q); \, \nu\, \rangle = -e f_V Q_V
m_V \, g^{\mu \nu} , \label{eq:014}
\end{equation}
where $g^{\mu \nu}$ is the metric tensor, \ $q$ is the photon (meson)
four-momentum and $Q_V$ is the effective electric charge of the quarks in the
meson $V$:
\begin{eqnarray}
&&Q_\rho =\frac{1}{\sqrt{2}} \,, \quad  Q_\omega=\frac{1}{3
\sqrt{2}}\,, \quad Q_\phi=- \frac{1}{3}\,,\nonumber \\   &&
Q_{J/\psi} = Q_{\psi(2S)}=\ldots = \frac{2}{3}\,. \label{eq:015}
\end{eqnarray}
The decay constant of the neutral vector meson $f_V$ can be
extracted from electromagnetic decay width, using
 \begin{equation}
\Gamma_{V \to e^+ e^-} = \frac{4 \pi \alpha_{em}^2 }{3\,m_V} f_V^2
Q_V^2\,. \label{eq:016}
 \end{equation}
We will call this version VMD1. The vertex in Eq.~(\ref{eq:014})
follows from the transition Lagrangian
 \begin{equation}
{\cal L}_{\gamma V} = -e f_V Q_V m_V \, A^\mu  \, V_\mu \,.
\label{eq:017}
 \end{equation}
Another model (called hereafter VMD2) originates from
 \begin{equation}
{\cal L}_{\gamma V} = - \frac{e f_V Q_V}{2 m_V} \, F^{\mu \nu} \,
V_{\mu \nu}\, , \label{eq:018}
\end{equation}
where $V_{\mu \nu} \equiv \partial_\mu V_\nu -
\partial_\nu V_\mu$ and $F^{\mu \nu} \equiv \partial^\mu A^\nu -
\partial^\nu A^\mu$. An advantage of the Lagrangian (\ref{eq:018}) is its explicit gauge-invariant
form.  Eq.~(\ref{eq:018}) gives rise to the $ V \, \gamma$
transition vertex
\begin{equation}
\langle \,\gamma (q); \, \mu | V (q); \, \nu  \,\rangle = -\frac{e
f_V Q_V }{m_V} \, (q^2 g^{\mu \nu} - q^\mu q^\nu )\,,
\label{eq:019}
\end{equation}
The term $\propto q^\mu q^\nu $ does not contribute being
contracted with the leptonic current $e \bar{u}(q_-) \gamma_\mu
v(q_+)$, and vertex (\ref{eq:019}) is suppressed compared to
(\ref{eq:014}) at small $q^2 \ll m_V^2$, i.e. in the region far
from the vector-meson mass shell. Of course on the mass shell,
$q^2 = m_V^2$, the VMD2 and VMD1 are equivalent.

These two versions of VMD model have been discussed earlier in
Refs.~\cite{Klingl:1996, O'Connell:1997}. Note also that VMD2
vertex follows from the Resonance Chiral Theory \cite{Ecker:1989}
and has been applied~\cite{Eidelman:2010} when studying the
reaction $ e^+ e^- \to \pi^0 \pi^0 (\eta) \gamma$.

Parameters of vector resonances are collected in Table~\ref{tab:param1}.
\begin{table*}[th]
\caption{Mass, total width, leptonic decay width and coupling
$f_V$ of vector mesons~\cite{PDG:2010} (experimental uncertainties
are not shown).}
\begin{center}
\begin{tabular}{c c c c c}
\hline $V$ & $m_V{\rm (MeV)}$ &$\Gamma_V{\rm (MeV)}$&$\Gamma_{V\to e^+\,e^-}{\rm (keV)}$&$f_V{\rm (MeV)}$  \\
\hline
$\rho^0$   & $775.49$ & $149.1$ & $7.04$ & $221.2$  \\
$\omega$   &$782.65$  & $8.49$ & $0.60$ & $194.7$   \\
$\phi$     &$1019.455$  & $4.26$ & $1.27$ & $228.6$  \\
$J/\psi$ & $3096.916$ & $0.0929$ & $5.55$ & $416.4$ \\
$\psi(2S)$ & $3686.09$ & $0.304$ & $2.35$ & $295.6$ \\
$\psi(3770)$ & $3772.92$ & $27.3$ & $0.265$ & $100.4$ \\
$\psi(4040)$ & $4039$ & $80$ & $0.86$ & $187.2$ \\
$\psi(4160)$ & $4153$ & $103$ & $0.83$ & $186.5$ \\
$\psi(4415)$ & $4421$ & $62$ & $0.58$ & $160.8$ \\
\hline
\end{tabular}
\end{center}
\label{tab:param1}
\end{table*}

The nonresonant amplitudes are calculated in the NFA, with the
short-distance NNLO Wilson coefficients, and nonperturbative $B
\to K^*$ transition form factors.

Then the total amplitudes including nonresonant and resonant parts
take the form
\begin{eqnarray} A_{0
L,R}& = &
\frac{1}{2\,\hat{m}_{K^*}\sqrt{\hat{q}^2}}\Biggl(C_0(q^2)\Bigl(C_{9V}^{\rm
eff} \mp C_{10A} \nonumber \\ &&+2\hat{m}_b\left(C_{7\gamma}^{\rm
eff}-C_{7\gamma}^{\prime\,\rm eff}\right)\kappa_0(q^2)\Bigr)
\nonumber \\
  && + 8\pi^2\sum_{V}C_V D_V^{-1}(\hat{q}^2)
\Bigl(\left(1-\hat{q}^2-\hat{m}_{K^*}^2\right)S_1^V\nonumber \\
&&+\hat{\lambda}\frac{S_2^V}{2}\Bigr)\Biggr)\,, \label{eq:020}
\end{eqnarray}
\begin{eqnarray}
A_{\| L,R}&=&-\sqrt{2}\Biggl(C_{\|}(q^2)\Bigl(C_{9V}^{\rm eff} \mp
C_{10A}\nonumber \\
&&+2\frac{\hat{m}_b}{\hat{q}^2}\left(C_{7\gamma}^{\rm
eff}-C_{7\gamma}^{\prime\,\rm
eff}\right)\kappa_{\|}(q^2)\Bigr)\nonumber \\
&&+8\pi^2\sum_{V}C_V
D_V^{-1}(\hat{q}^2)\,S_1^V\Biggr)\,,\label{eq:021}
\end{eqnarray}
\begin{eqnarray}
A_{\perp
L,R}&=&\sqrt{2\hat{\lambda}}\Biggl(C_{\perp}(q^2)\Bigl(C_{9V}^{\rm
eff} \mp
C_{10A}\nonumber \\
&&+2\frac{\hat{m}_b}{\hat{q}^2}\left(C_{7\gamma}^{\rm
eff}+C_{7\gamma}^{\prime\,\rm
eff}\right)\kappa_{\perp}(q^2)\Bigr)\nonumber \\
&&+4\pi^2\sum_{V}C_V
D_V^{-1}(\hat{q}^2)\,S_3^V\Biggr)\,,\label{eq:022}
\end{eqnarray}
\begin{equation}
A_t=-2\sqrt{\frac{\hat{\lambda}}{\hat{q}^2}}\,C_{10A}\,A_0(q^2)\,,\label{eq:023}
\end{equation}
where
\[ D_V(\hat{q}^2) = \hat{q}^2 - \hat{m}_V^2 +
i\hat{m}_V \hat{\Gamma}_V (\hat{q}^2) \] is the usual Breit-Wigner
function for the $V$ meson resonance shape with the
energy-dependent width ${\Gamma}_V ({q}^2)$ \ [$\hat{\Gamma}_V
(\hat{q}^2) = {\Gamma}_V ({q}^2) /m_B $], $\hat{m}_V\equiv
m_V/m_B$, $\hat{\Gamma}_V\equiv \Gamma_V/m_B$, $m_V(\Gamma_V)$ is
the mass (width) of a $V$ meson and the form factors enter as
\begin{eqnarray}
C_0(q^2)&=&(1-\hat{q}^2-\hat{m}_{K^*}^2)(1+\hat{m}_{K^*})
A_1(q^2)\nonumber \\ &&-\hat{\lambda}
\frac{A_2(q^2)}{1+\hat{m}_{K^*}} \label{eq:024},
\end{eqnarray}
\begin{equation}
C_{\|}(q^2)=(1+\hat{m}_{K^*})A_1(q^2) \label{eq:025},
\end{equation}
\begin{equation}
C_{\perp}(q^2)=\frac{V(q^2)}{1+\hat{m}_{K^*}} \label{eq:026},
\end{equation}
\begin{eqnarray}
\kappa_0(q^2)&\equiv&
\Bigl((1-\hat{q}^2+3\hat{m}_{K^*}^2)(1+\hat{m}_{K^*})T_2(q^2)\nonumber
\\&&-\frac{
\hat{\lambda}}{1-\hat{m}_{K^*}}T_3(q^2)\Bigr)\Bigl((1-\hat{q}^2-\hat{m}_{K^*}^2)\nonumber
\\&&\times(1+\hat{m}_{K^*})^2 A_1(q^2)-\hat{\lambda}\,
A_2(q^2)\Bigr)^{-1} \label{eq:027},
\end{eqnarray}
\begin{equation}
\kappa_{\|}(q^2)\equiv \frac{T_2(q^2)}{A_1(q^2)}(1-\hat{m}_{K^*}),
\label{eq:028}
\end{equation}
\begin{equation}
\kappa_{\perp}(q^2)\equiv \frac{T_1(q^2)}{V(q^2)}(1+\hat{m}_{K^*})
. \label{eq:029}
\end{equation}
In the above formulas the definition $\hat{m}_b\equiv
\overline{m}_b(\mu)/m_B$, $\hat{m}_s\equiv
\overline{m}_s(\mu)/m_B$ are used, and $\overline{m}_b(\mu)$
[$\overline{m}_s(\mu)$] is the running bottom (strange) quark mass
in the $\overline{\rm MS}$ scheme at the scale $\mu$.

The SM Wilson coefficients have been obtained in
\cite{Altmannshofer:2009} at the scale $\mu=4.8$ GeV to NNLO
accuracy and equal
\[
C_{7\gamma}^{\rm eff}(\mu)= -0.304\, , \; C_{9{\rm
V}}(\mu)=4.211\, , \; C_{10{\rm A}}(\mu)=-4.103 \, ,\]
$C_{9V}^{\rm eff} = C_{9V} +Y(q^2)$, where $Y(q^2)$ is quark-loop
function given in Ref.~\cite{Beneke:2001a}. Note that in the
framework of the SM  \ $\overline{m}_b(\mu) \,
C_{7\gamma}^{\prime\,\rm eff}= \overline{m}_s (\mu)
\,C_{7\gamma}^{\rm eff}$.

Further, $A_0(q^2)$, $A_1(q^2)$, $A_2(q^2)$, $V(q^2)$, $T_1(q^2)$,
$T_2(q^2)$, $T_3(q^2)$ are the $B \to K^*$ transition form
factors. In the numerical estimations, we use the form factors
from the LCSR calculation \cite{Ball:2005}.

In Eqs.~(\ref{eq:020})-(\ref{eq:022}), $S_i^V$ \ ($i=1,2,3$) are
the invariant amplitudes of the decay $B_d^0\to  K^{*0}\,V$. These
amplitudes are calculated in~\ref{subsec:vector mesons}. The
coefficients $C_V$ in the resonant contribution are
\[C_V= \frac{Q_V m_V f_V}{q^2}\,\left(\rm{VMD1}\right), \quad
 C_V= \frac{Q_V f_V}{m_V}\,\left(\rm{VMD2}\right)\, .\]

The energy-dependent widths of light vector resonances $\rho$,
$\omega$ and $\phi$ are chosen as in Ref.~\cite{Korchin:2010}. The
up-dated branching ratios for resonances decays to different
channels are taken from \cite{PDG:2010}. For the $c \bar{c}$
resonances $J/\psi$, \ $\psi (2S)$, $\ldots$ we take the constant
widths.

In order to calculate the resonant contribution to the amplitude of the ${\bar
B}_d^0\to {\bar K}^{*0}\,\mu^+\,\mu^-$ decay, one has to know the amplitudes of
the decays ${\bar B}_d^0\to {\bar K}^{*0}\, \rho$, \ ${\bar B}_d^0\to {\bar
K}^{*0}\,\omega, \, {\bar B}_d^0\to {\bar K}^{*0}\,\phi, \, {\bar B}_d^0\to
{\bar K}^{*0}\,J/\psi, \, {\bar B}_d^0\to {\bar K}^{*0}\,\psi(2S)$. The
information on the ${\bar B}_d^0\to {\bar K}^{*0}\,\phi, \, {\bar B}_d^0\to
{\bar K}^{*0}\,J/\psi, \, {\bar B}_d^0\to {\bar K}^{*0}\,\psi(2S) $ decays can
be taken from experiment~\cite{PDG:2010}. For the light resonances $\rho$ and $
\omega$ we use the theoretical prediction~\cite{Chen:2006} for the decay
amplitudes. At the same time, we are not aware of a similar prediction for the
higher $c \bar{c}$ resonances, such as $\psi(3770)$ and so on, and therefore
do not include these resonances.

The parameters of the model are indicated in
Table~\ref{tab:param2}.

\begin{table}[t]
\caption{The numerical input used in our analysis.}
\label{tab:param2}
\begin{center}
\begin{tabular}{ll}
\hline
$|V_{tb}V_{ts}^*|=0.04026$ & $G_F=1.16637\times 10^{-5}\, {\rm GeV^{-2}}$\\
$\mu=m_b=4.8\, {\rm GeV}$   & $\alpha_{\rm em}=1/137.036$\\
$m_c=1.4\, {\rm GeV}$       &$m_B=5.27950\, {\rm GeV}$ \\
$\overline{m}_b(\mu)=4.14\, {\rm GeV}$     &$\tau_B=1.525\, {\rm ps}$ \\
$\overline{m}_s(\mu)=0.079\, {\rm GeV}$     &$m_{K^*}=0.89594\, {\rm GeV}$\\
\hline
\end{tabular}
\end{center}
\end{table}

\section{ \label{sec:results}
 Results of the calculation for the ${\bar B}_d^0\to {\bar
K}^{*0}\, \mu^+ \, \mu^-$ decay }

\begin{figure*}
\begin{center}
\includegraphics[width=.42\textwidth]{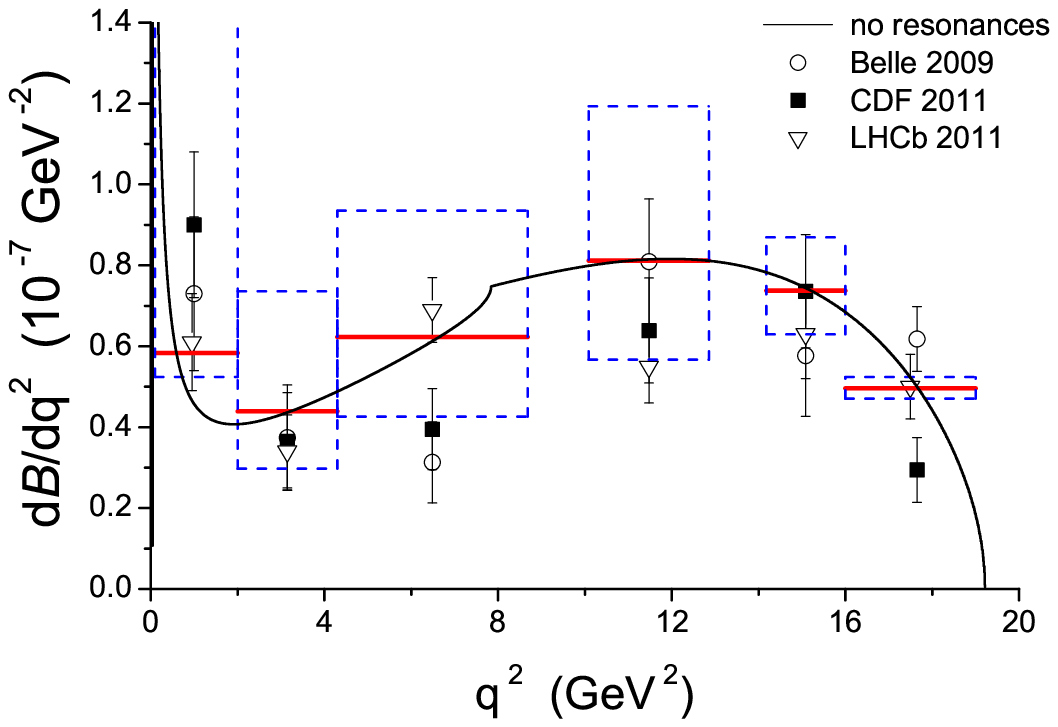}
\includegraphics[width=.42\textwidth]{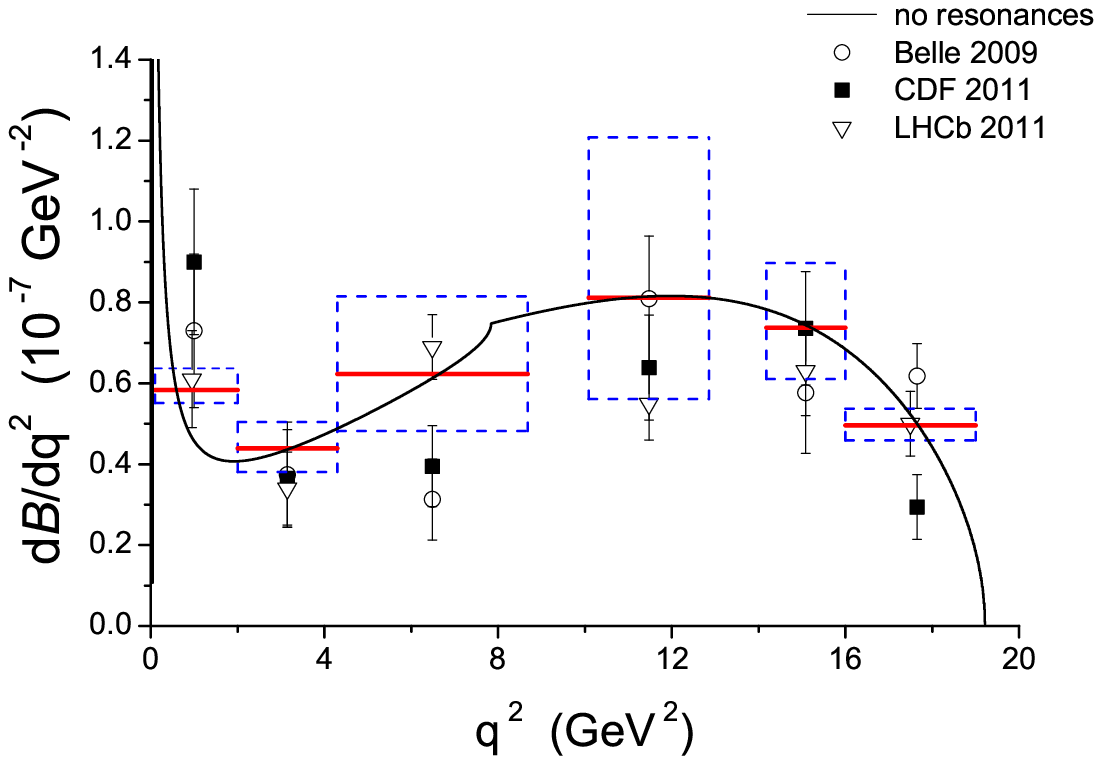}
\caption{Differential branching ratio as a function of $q^2$. Solid lines
correspond to calculation in the SM without resonances, and the horizontal bars
(red online) are the corresponding binned (averaged over the experimental bins)
results. In the first bin calculation is performed for $0.1 \, {\rm GeV}^2 <
q^2 < 2.0 \, {\rm GeV}^2$.
The dashed rectangles (blue online) show contribution of the intermediate
resonances: left (right) panel - in the VMD1 (VMD2) model. The uncertainty of
the resonance contribution is related to the unknown phase $\delta_0^{V}$ of
the resonance amplitudes (see the text). The upper side of the resonance
rectangle (left side) is located at 3.54 and is not shown. The form factors are
taken from~\cite{Ball:2005}. The data Belle (KEKB)~\cite{Belle:2009}, CDF
(Tevatron)~\cite{CDF:2011, CDF:2011a} and LHCb~\cite{LHCb:2011} are shown by
the circles, filled boxes and triangles respectively (the horizontal error bars
are not shown - they are given in Table~\ref{tab:bins}). } \label{fig:Br}
\end{center}
\end{figure*}

\begin{figure*}
\begin{center}
\includegraphics[width=.42\textwidth]{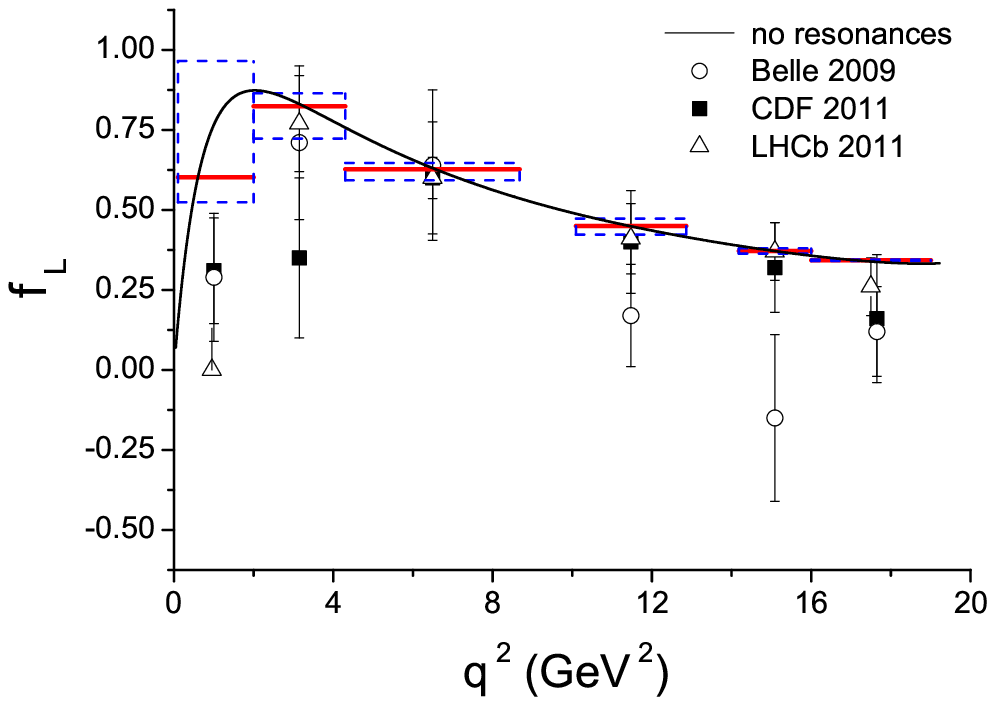}
\includegraphics[width=.42\textwidth]{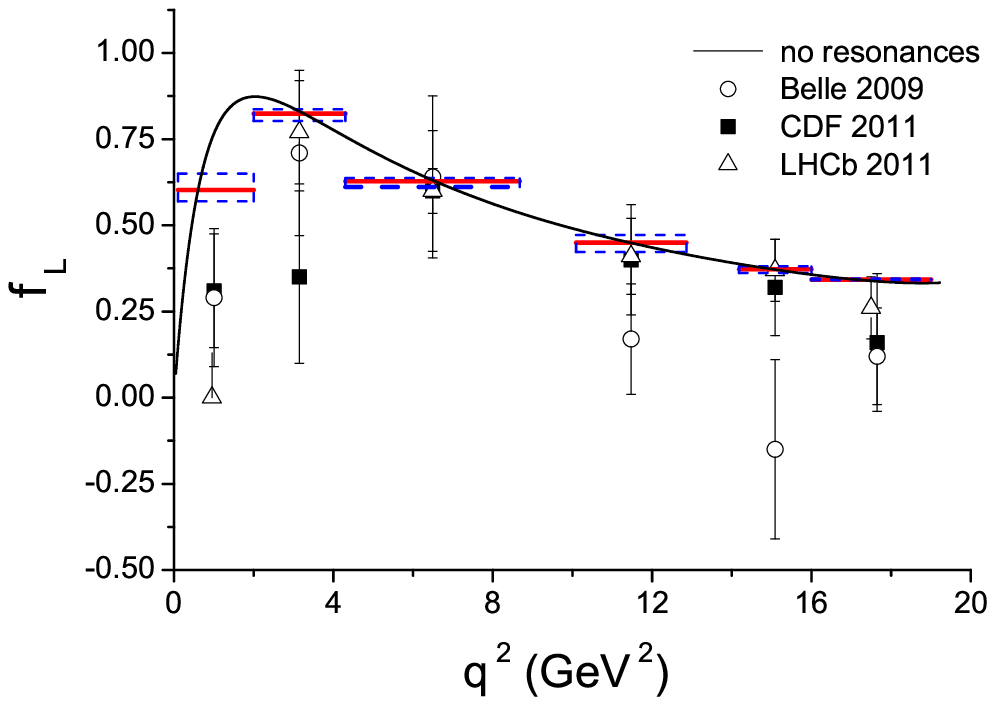}
\includegraphics[width=.42\textwidth]{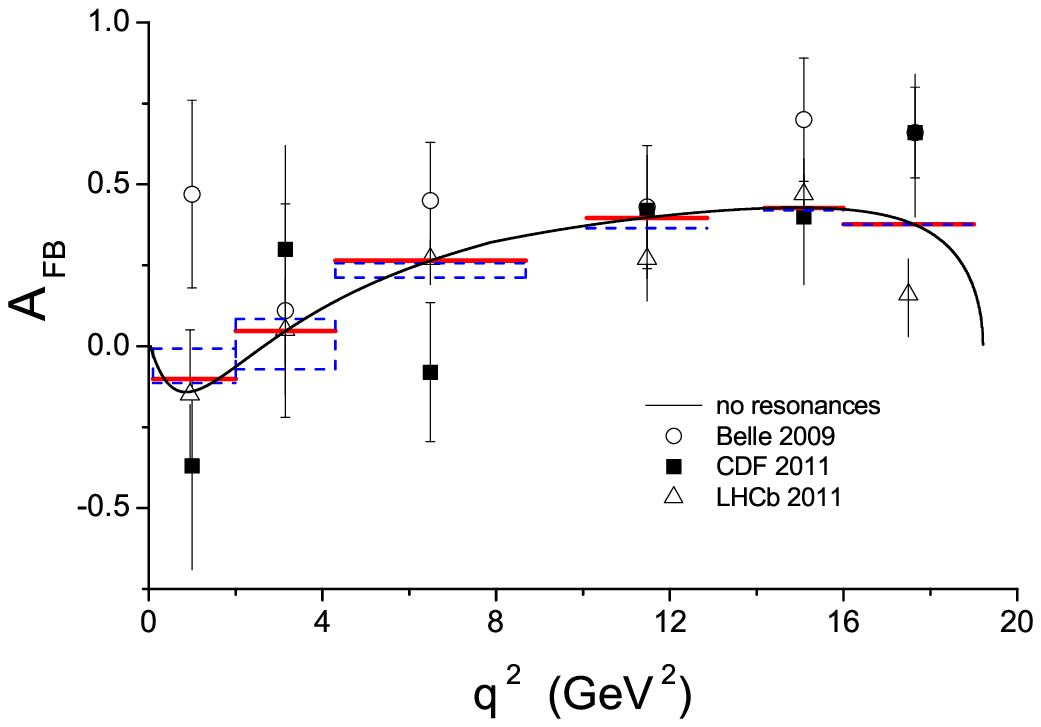}
\includegraphics[width=.42\textwidth]{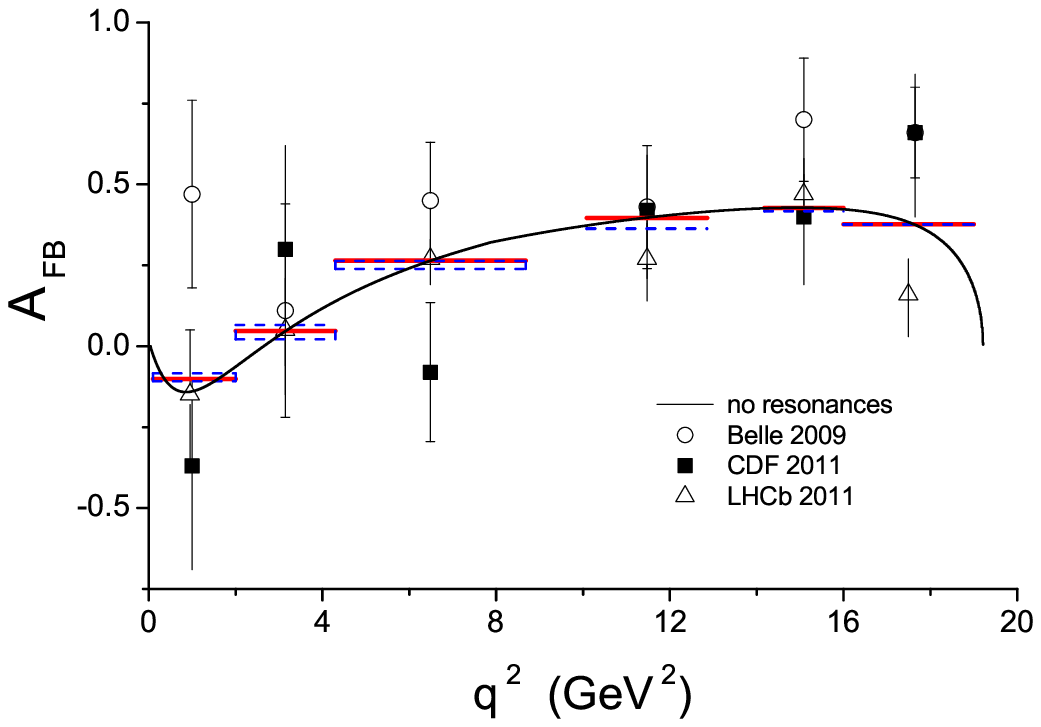}
\caption{Longitudinal polarization fraction of $K^*$ meson (upper
panel) and forward-backward asymmetry $A_{\rm FB}$ (lower panel)
as functions of $q^2$. Due to the choice of reference frame in
Fig.~\ref{fig1}, the forward-backward asymmetry $A_{\rm FB}$ in
Refs.~\cite{Belle:2009, CDF:2011, CDF:2011a, LHCb:2011} is related
to asymmetry in Eq.~(\ref{eq:009}) via $A_{\rm FB} = - d{\bar
A}_{\rm FB}^{(\mu)}/d\hat{q}^2$. The other notation is the same as
in Fig.~\ref{fig:Br}.} \label{fig:fL-AFB}
\end{center}
\end{figure*}

In Figs.~\ref{fig:Br} and \ref{fig:fL-AFB} we present results for
the dependence of several observables in the ${\bar B}_d^0\to
{\bar K}^{*0}\, \mu^+\, \mu^-$ decay on the dimuon invariant mass
squared. The interval of $q^2$ is taken from $4 m_{\mu}^2$ to
$q_{max}^2 = (m_B- m_{K^*})^2 \approx 19.22$ GeV$^2$.
\begin{table}[tbh]
\caption{Experimental bins in $q^2$ used in the data analyses
\cite{LHCb:2011} (the first column), \cite{Belle:2009, CDF:2011,
CDF:2011a} (the second column) and in our calculation.}
\begin{center}
\begin{tabular}{c | c }
\hline
$q^2$ (GeV$^2$) &  $q^2$ (GeV$^2$) \\
\hline
0.10 - 2.00 &  0.00 - 2.00 \\
2.00 - 4.30 & 2.00 - 4.30 \\
4.30 - 8.68 & 4.30 - 8.68 \\
10.09 - 12.86 & 10.09 - 12.86 \\
14.18 - 16.00 & 14.18 - 16.00 \\
16.00 - 19.00 &  16.00 - $q_{max}^2$ \\
\hline
\end{tabular}
\end{center}
\label{tab:bins}
\end{table}

In order to adequately compare results of calculations with
experiments from Belle (KEKB)~\cite{Belle:2009}, CDF
(Tevatron)~\cite{CDF:2011, CDF:2011a} and LHCb~\cite{LHCb:2011} we
show the binned predictions, i.e. the calculated values averaged
over the experimental bins. These bins are indicated in
Table~\ref{tab:bins}.

The solid line demonstrate the prediction of the SM without
resonances for the $q^2$-dependence of all observables; the
horizontal bars are the corresponding binned results.

The dashed rectangles in Figs.~\ref{fig:Br} and \ref{fig:fL-AFB} show possible
contributions of the intermediate vector resonances in each $q^2$-bin. To
obtain these rectangles the phases $\delta_0^{V} ={\rm arg}(A_0^V)$ of the $B
\to K^* V$ decay amplitudes with the zero helicity for $V = \rho, \, \omega, \,
\phi, \, J/ \psi, \, \psi(2S) $ have been varied from 0 to $2\pi$. The upper
(lower) side of the rectangles corresponds to the maximal (minimal) value of
the branching ratio in each bin. The dominant dependence on the phase
$\delta_0^{V}$ comes from the $c \bar{c}$ resonances $J/\psi$ and $\psi(2S)$,
while the phase dependence for the $\rho$, $\omega$ and $\phi$ is of minor
importance. This is explained by the much bigger contribution of the $c
\bar{c}$ resonances compared to the light ones.

Predictions of the VMD1 and VMD2 models
(Section~\ref{subsec:resonances}) differ at relatively small
$q^2$. This is due to the fact that at small $q^2$ the high-lying
resonances $J/ \psi$ and $\psi(2S)$ are far off their mass shells
and in these kinematical conditions the VMD1 and VMD2 models give
different results. On the contrary, in the high $q^2$ region,
close to the mass shells for the $c \bar{c}$ resonances, the
results with VMD1 and VMD2 do not differ.

In the experimental analyses certain cuts in $q^2$ are applied in order to
suppress the charmonia contributions (the so-called charmonia veto).
Correspondingly in Figs.~\ref{fig:Br} and \ref{fig:fL-AFB} we also use these
cuts. As it is seen from Fig.~\ref{fig:Br}, the charmonia resonances may
contribute to the differential branching far beyond their pole positions and
beyond the charmonia veto, although their contribution crucially depends on the
values of the zero-helicity phases $\delta_0^{J/\psi}$ and
$\delta_0^{\psi(2S)}$. The observed in the VMD1 model sizable contribution of
the $c \bar{c}$-resonances at small values of $q^2$ occurs at certain values of
these phases. Of course there exist values of $\delta_0^{J/\psi}$ and
$\delta_0^{\psi(2S)}$ at which resonance contribution is small.

Comparing these results with results of
Refs.~\cite{Ali:2000,Ligeti:1996,Ligeti:1998} we note that in these papers the
amplitudes of the decays $B \to K^* \, J/\psi$ and  $B \to K^* \, \psi(2S) $
are calculated in framework of the NFA which implies zero phases. In contrast,
our calculation is free from this limitation since the amplitudes are taken
from experiment and thus the phases $\delta_0^V$ for all resonances $V$ can be
chosen arbitrary and independent.

Note that long-distance effects due to the nonfactorizable soft-gluon emission from the $c$-quarks have been included in Ref.~\cite{Khodjamirian:2010}. Authors have shown that these effects lead to a modification, depending on the $K^*$ polarization, of the Wilson coefficient $C_{9V}(\mu)$. The form factors and the Wilson coefficients in~\cite{Khodjamirian:2010} are different from the corresponding quantities in our calculation. In framework of this approach
we estimated the differential branching ratio in the region $1.0 \, {\rm GeV}^2 \, <  q^2  < \, 9.0 \, {\rm GeV}^2$, using the Wilson coefficients and form factors of our work. Results of this calculation along with results of the present work are shown in Table~\ref{tab:comparison}.
\begin{table}[tbh]
\caption{Differential branching ratio averaged over the two
experimental bins. Results in our approach with resonances (VMD1 and VMD2) show variation of the branching ratio depending on the phases $\delta_0^V$. Values in the last row are obtained following approach~\cite{Khodjamirian:2010} for the central values of all
parameters in Eq.~(70) from~\cite{Khodjamirian:2010}
without taking into account uncertainties.
}
\begin{tabular}{l|cc}
\hline
 & \multicolumn{2}{c}{$ d B/ dq^2 \, \,  (10^{-7} \ \rm{GeV}^{-2})$}   \\
\noalign{\smallskip}
\hline
$q^2 \, \,  (\rm{GeV}^2)$ & 2.00 - 4.30 & 4.30 - 8.68 \\
\hline
SM without resonances  &  0.44 & 0.63 \\
VMD1  &  0.30 - 0.74 & 0.43 - 0.94 \\
VMD2  &  0.38 - 0.50 & 0.48 - 0.82 \\
Ref.~\cite{Khodjamirian:2010} & 0.44 & 0.60 \\
\hline
\end{tabular}
\label{tab:comparison}
\end{table}

As is seen, results of both approaches are comparable,
at least in these intervals of $q^2$ and for a certain choice of phases
$\delta_0^V$. Besides, our calculation in the SM without resonances in framework of the NFA turns out to be relatively close to the calculation following approach of Ref.~\cite{Khodjamirian:2010}.

As expected, the observables which are the ratios of the bilinear combinations
of the amplitudes, such as $f_L$ or $A_{\rm FB}$, are less sensitive to the
phases of the resonance contribution (see Fig.~\ref{fig:fL-AFB}). Moreover, the
latter is generally small independently of $\delta_0^V$, apart from the region
of relatively small $q^2 \lesssim 4$ GeV$^2$.

It should be noted that for the inclusive decay $B \to X_s \, e^+ e^- $, in
Ref.~\cite{Buchalla:1998} soft-gluon emission from the charm loop has been
considered and long-distance corrections of ${\cal O} (\Lambda_{QCD}^2 /m_c^2)$
to the differential branching and forward-backward asymmetry are shown to be
small in the region away from the $c \bar{c}$ resonances. Although this
inclusive process is amenable to clean theoretical description for $1$ GeV$^2$
$ < q^2 < 6 $ GeV$^2$ and $q^2 > 14.4$ GeV$^2$~\cite{Hurth:2010}, the
experimental information on differential branching ratio is still rather
limited~\cite{Babar:2004,Belle:2005}. The situation is not likely to improve in
the nearest future.
At the same time for the exclusive decays $ B \to K^*(\to K\pi)\,\ell^+
\,\ell^-$ a high statistics is expected at LHCb, and this will allow for
measurement of various observables in these processes. Therefore the exclusive
decays have quite a good potential for tests of the SM and search for effects
beyond the SM.

\section{ \label{sec:conclusions} Conclusions}

The rare FCNC decay ${\bar B}_d^0 \to {\bar K}^{*0} \, (\to K^{-}\, \pi^+) \,
\mu ^+\, \mu^- $ has been studied in the whole region of muon-antimuon
invariant masses. We performed calculations of the differential branching
ratio, polarization fraction $f_L$ of the $K^*$ meson and forward-backward
asymmetry $A_{{\rm FB}}$ for the four-body decay ${\bar B}_d^0 \to K^{-}\,
\pi^+ \, \mu^+\,\mu^- $.
Main emphasis in our study is placed on contribution of the intermediate vector
resonances in the process $\bar{B}_d^0 \to \bar{K}^{*0} \, (\to K^{-}\, \pi^+)
\, V $ with $V = \rho(770), \, \omega(782), \, \phi(1020), \,$  $J/\psi, \,
\psi(2S)$, decaying into the $\mu^+ \mu^-$ pair. Various aspects of theoretical
treatment of this long-distance contribution have been investigated.

One aspect is the choice of VMD model describing the $V \, \gamma$ transition.
We applied two different versions called VMD1 and VMD2. In particular, VMD2
model explicitly obeys gauge invariance and the corresponding $V \, \gamma$
vertex is suppressed compared to the $V \, \gamma$ vertex in the VMD1 model at
small values of the photon invariant mass, far from the vector-meson mass
shell. This turns out to be important when calculating the contribution of the
high-lying $J/\psi$ and $\psi(2S)$ resonances to the branching ratio at $q^2
\ll m_V^2$.

Another aspect is the structure of the vertex $\bar{B}_d^0 \to \bar{K}^{*0} \,
V $ with the off-shell vector meson. We used an off-mass-shell extension of the
helicity amplitudes which describe production of the on-shell vector mesons
$V$. For the latter amplitudes all the experimentally available information is
used, and otherwise theoretical predictions.

The calculations are compared with recent data~\cite{Belle:2009,CDF:2011,
CDF:2011a, LHCb:2011} for the $q^2$-dependence of the differential branching
ratio, longitudinal polarization fraction of the $K^*$ meson and
forward-backward asymmetry. In these analyses the cuts around the pole
positions of the charmonia resonances are applied to suppress these
contributions (charmonia veto). Our calculation shows that the intermediate
resonances may considerably modify the branching ratio, calculated in the SM
without resonances in framework of the NFA, even in the $q^2$-region located
far from these cuts. Main contribution comes from the $c \bar{c}$ resonances
$J/\psi$ and $\psi(2S)$.
This conclusion, however, crucially depends on values of the unknown
zero-helicity phases $\delta_0^{J/\psi}$ and $\delta_0^{\psi(2S)}$ for the
decays $B^0 \to K^{*0} \, J/\psi$ and $B^0 \to K^{*0} \, \psi(2S)$
respectively. In view of these results one should keep in mind that the
resonance contribution may imitate effects of new physics, the search for which
is one of the goals of current and planned experiments on BaBar, CDF, Belle and
LHCb.

As for the observables $f_L$ and $A_{\rm FB}$, these are less sensitive to the
phases of the resonance contribution and moreover the latter is generally small
independently of $\delta_0^V$ (apart from the region of relatively small $q^2
\lesssim 4$ GeV$^2$ in the VMD1 model). Apparently data for these observables
are more appropriate for adequate comparison with the prediction of the SM
without resonances.

\appendix

\section{\label{subsec:vector mesons} Amplitudes of $B \to K^* V$
decays }

An important ingredient of the resonant contribution is amplitude
of the decay of $B$ meson into two vector mesons, \ $B(p) \to
V_1(q,\epsilon_1) + V_2(k,\epsilon_2)$, with on-mass-shell meson
$V_2$ ($k^2 = m_2^2$) and off-mass-shell meson $V_1$ ($q^2 \ne
m_1^2$).
\begin{table*}[tbh]
\caption{Branching ratio \cite{PDG:2010}, and decay amplitudes for
${ B}_d^0\to {K}^{*0}\,\rho^0$ \cite{Chen:2006}, ${B}_d^0\to {
K}^{*0}\,\omega$ \cite{Chen:2006} and ${ B}_d^0\to {
K}^{*0}\,\phi$, ${ B}_d^0\to { K}^{*0}\,J/\psi$, ${ B}_d^0\to {
K}^{*0}\,\psi(2S)$ \cite{PDG:2010}.}
\begin{center}
\begin{tabular}{c c c c c c}
\hline
 $V$ & $\rho^0$ &$\omega$&$\phi$&$J/\psi$&$\psi(2S)$ \\
\hline $10^6{\rm BR}({ B}_d^0\to { K}^{*0}\,V)$&$3.4$&$2.0$&$9.8$&$1330$&$610$\\
$|h_0^V|^2$       & $0.70$ & $0.75$ & $0.480$&$0.570$&$0.46$ \\
$|h_\perp^V|^2$   &$0.14$  & $0.12$ & $0.24$&$0.219$&$0.30$\\
${\rm arg}(h_\|^V/h_0^V)$ (rad)&$1.17$  & $1.79$ & $2.40$&$-2.86$&$-2.8$\\
${\rm arg}(h_\perp^V/h_0^V)$ (rad)&$1.17$& $1.82$ & $2.39$&$3.01$&$2.8$ \\
$10^4|S_1^V|$ &$1.17$&$0.81$&$2.66$&$33.64$&$28.86$ \\
$10^4|S_2^V|$ &$2.65$&$1.67$&$5.20$&$42.49$&$52.65$ \\
$10^4|S_3^V|$ &$2.31$&$1.64$&$5.28$&$115.28$&$153.00$ \\
$\delta_1^V-\delta_0^V$ (rad)&$1.17$  & $1.79$ & $2.40$&$-2.86$&$-2.8$\\
$\delta_2^V-\delta_0^V$ (rad)&$-2.11$  & $-1.53$ & $-0.84$&$0.90$&$1.62$\\
$\delta_3^V-\delta_0^V$ (rad)&$1.17$& $1.82$ & $2.39$&$3.01$&$2.8$ \\
\hline
\end{tabular}
\end{center}
 \label{tab:ampl}
\end{table*}
For the case of two on-mass-shell final mesons one can write the
amplitude in the form \cite{Valencia:1989}
\begin{eqnarray}
{\cal M} &=&\frac{G_F\,m_B^3}{\sqrt{2}}|V_{\rm CKM}|\Bigl(S_1\,
g_{\mu\nu} + \frac{S_2}{m_B^2}\, p_\mu p_\nu \nonumber \\&&- i
\frac{S_3}{m_B^2} \, \varepsilon_{\mu\nu\alpha\beta}\, q^\alpha
k^\beta \Bigr)\epsilon_1^{\mu *}\epsilon_2^{\nu *}\label{eq:0A1}
\end{eqnarray}
in terms of three invariant amplitudes $S_1$, $S_2$ and $S_3$,
$V_{\rm CKM}$ is a CKM factor. The quantities $S_1$, $S_2$ and
$S_3$ may be complex and involve two types of phases,
$CP$-conserving strong phases and $CP$-violating weak phases. In
general, the invariant amplitudes are a sum of several interfering
amplitudes, $S_{1j}$, $S_{2j}$ and $S_{3j}$, respectively. Then
the phase structure of $S_1$, $S_2$ and $S_3$ is:
\begin{equation}\label{eq:0A2}
S_k=\sum_{j}|S_{kj}|\,e^{i \varphi_{kj}}e^{i \delta_{kj}}\, \qquad
\qquad (k=1,2,3 ) \,,
\end{equation}
where $\varphi_{1j}$, $\varphi_{2j}$, and $\varphi_{3j}$ are the
$CP$-violating weak phases and $\delta_{1j}$, $\delta_{2j}$, and
$\delta_{3j}$ are the $CP$-conserving strong phases.

Using $CPT$ invariance, we can represent the matrix element for
the charge-conjugate decay $\bar B(p) \to \bar
V_1(q,\epsilon_1)\,\bar V_2(k,\epsilon_2)$ as
\begin{eqnarray}\label{eq:0A3}
\overline{{\cal M}}&=&\frac{G_F\,m_B^3}{\sqrt{2}}|V^*_{\rm
CKM}|\Bigl(\bar S_1\, g_{\mu\nu} + \frac{\bar S_2}{m_B^2}\, p_\mu
p_\nu \nonumber \\&&+ i \frac{\bar S_3}{m_B^2} \,
\varepsilon_{\mu\nu\alpha\beta}\, q^\alpha k^\beta
\Bigr)\epsilon_1^{\mu *}\epsilon_2^{\nu *}\,,
\end{eqnarray}
where $\bar S_1$, $\bar S_2$, and $\bar S_3$ can be derived from
$S_1$, $S_2$, and $S_3$ by reversing the sign of the
$CP$-violating phase. Note that if the $B\to V_1\,V_2$ decay is
invariant under the $CP$ symmetry, then $\bar S_1=S_1$, $\bar
S_2=S_2$, and $\bar S_3=S_3$. On the other hand, if all
$CP$-conserving phases of invariant amplitudes are equal to zero,
then $\bar S_1=S^*_1$, $\bar S_2=S^*_2$, and $\bar S_3=S^*_3$.

The helicity amplitudes in terms of three invariant amplitudes,
$S_1$, $S_2$, and $S_3$ are:
\begin{eqnarray}\label{eq:0A4}
H_\lambda&\equiv&\Bigl(S_1\, g_{\mu\nu} + \frac{S_2}{m_B^2}\,
p_\mu p_\nu \nonumber \\ &&- i \frac{S_3}{m_B^2} \,
\varepsilon_{\mu\nu\alpha\beta}\, q^\alpha k^\beta
\Bigr)\epsilon_1^{\mu *}(\lambda)\epsilon_2^{\nu *}(\lambda)\,.
\end{eqnarray}

From the decomposition Eq.~(\ref{eq:0A4}) one finds the following
relations between the helicity amplitudes and the invariant
amplitudes $S_1$, $S_2$, $S_3$:
\begin{eqnarray} \label{eq:0A5}
H_0 &=& - \frac1 {2 \hat{m}_1 \hat{m}_2} \Bigl(
(1-\hat{m}_1^2-\hat{m}_2^2)  S_1 \nonumber \\&&+ \frac{S_2}{2}
\lambda(1,\hat{m}_1^2,\hat{m}_2^2) \Bigr), \nonumber \\
H_\pm &=& S_1 \pm \frac{S_3}{2}
\sqrt{\lambda(1,\hat{m}_1^2,\hat{m}_2^2)} ,
\end{eqnarray}
with $\lambda(1,\hat{m}_1^2,\hat{m}_2^2) \equiv (1-\hat{m}_1^2)^2
- 2\hat{m}_2^2(1+\hat{m}_1^2) +\hat{m}_2^4$ and $\hat{m}_{1(2)}
\equiv m_{1(2)}/m_B$.

Note that the polarized decay amplitudes can be expressed in
several different but equivalent bases. For example, the helicity
amplitudes can be related to the spin amplitudes in the
transversity basis $\left(A_0\,,A_\|\,,A_\perp \right)$ defined in
terms of the linear polarization of the vector mesons via:
\[A_0 = H_0\,,\quad A_\parallel = \frac{H_+ + H_-}{\sqrt{2}}\,,
\quad A_\perp = \frac{H_+ - H_-}{\sqrt{2}}  \,,\]
$A_0$, $A_\|$, $A_\perp$ are related to $S_1$, $S_2$ and $S_3$ of
Eq.~(\ref{eq:0A1}) via
\begin{eqnarray}\label{eq:0A6}
A_0 &=& -\frac1 {2 \hat{m}_1 \hat{m}_2} \Bigl(
(1-\hat{m}_1^2-\hat{m}_2^2)  S_1 \nonumber \\&&+ \frac{S_2}{2}
\lambda(1,\hat{m}_1^2,\hat{m}_2^2) \Bigr), \nonumber \\
A_\parallel &=& \sqrt{2} \, S_1\,,\quad A_\perp =
\sqrt{\frac{\lambda(1,\hat{m}_1^2,\hat{m}_2^2)}{2}}\, S_3\,.
\end{eqnarray}
The amplitude $\bar A_\lambda$ ($\lambda=0\,,\|\,,\perp$) are
related to the invariant amplitudes of the $\bar B \to \bar
V_1\,\bar V_2$ decay by the formulas
\begin{eqnarray}\label{eq:0A7}
\bar A_0 &=& - \frac1 {2 \hat{m}_1 \hat{m}_2} \Bigl(
(1-\hat{m}_1^2-\hat{m}_2^2) \,\bar S_1 \nonumber \\&&+ \frac{\bar
S_2}{2}\, \lambda(1,\hat{m}_1^2,\hat{m}_2^2) \Bigr)\,, \nonumber
\\ \bar A_\parallel &=& \sqrt{2} \, \bar S_1\,, \quad \bar
A_\perp =-
\sqrt{\frac{\lambda(1,\hat{m}_1^2,\hat{m}_2^2)}{2}}\,\bar S_3\,.
\end{eqnarray}
If the $B\to V_1\,V_2$ decay is invariant under $CP$
transformation, then $\bar A_0=A_0$, $\bar A_\|=A_\|$, and $\bar
A_\perp=-A_\perp$.

The decay width is expresses as follows:
\begin{eqnarray} \label{eq:0A8}
\Gamma(B \to V_1 V_2)&=&
\frac{m_B\,\sqrt{\lambda(1,\hat{m}_1^2,\hat{m}_2^2)}}{16\pi}
\left(\frac{G_F m_B^2}{\sqrt{2}}\,|V_{\rm CKM}|\right)^2 \nonumber
\\ &&\times\left( |A_0|^2 + |A_\||^2 + |A_\perp|^2 \right) \,.
\end{eqnarray}

The matrix element for the $B_d^0\to  K^{*0}\,V$ decay, where
$V=\rho^0\,,\omega\,, \phi\,,J/\psi(1S)\,,\psi(2S)\,,\ldots$
mesons, we can represent as
\begin{eqnarray}\label{eq:0A9}
{\cal M}
&=&\frac{G_F\,m_B^3}{\sqrt{2}}|V^*_{tb}\,V_{ts}|\Bigl(S^V_1\,
g_{\mu\nu} + \frac{S^V_2}{m_B^2}\, p_\mu p_\nu \nonumber \\ &&- i
\frac{S^V_3}{m_B^2} \, \varepsilon_{\mu\nu\alpha\beta}\, q^\alpha
k^\beta \Bigr)\epsilon_1^{\mu *}\epsilon_2^{\nu *}.
\end{eqnarray}
Next, we define the normalized amplitudes:
\begin{eqnarray} \label{eq:0A10}
&&h_\lambda \equiv \frac{A_\lambda}{\sqrt{\sum_{\lambda^\prime}
|A_{\lambda^\prime}|^2}}\,, \nonumber \\ &&\sum_\lambda
|h_\lambda|^2 = 1\, \quad (\lambda, \lambda^\prime = 0,
\parallel, \perp)\,.
\end{eqnarray}
By putting $m_1 = m_V$, \ $m_2 = m_{K^*}$ and using
(\ref{eq:0A8}), (\ref{eq:0A10}) we obtain the relation between the
amplitudes $h_\lambda$ and $A_\lambda$ of the process under study
$B_d^0 \to K^{*0}\, V$ for any vector meson $V=\rho^0\,,\omega\,,
\phi\,,J/\psi(1S)\,,\psi(2S)\,,\ldots$:
\begin{eqnarray} \label{eq:0A11}
h_\lambda^V &=& \frac{G_F m_B^2}{4\sqrt{2}}|V^*_{tb}\,V_{ts}|
\sqrt{\frac{ m_B\,\tau_B}{\pi \, {\rm BR}(B_d^0 \to K^{*0}\, V) }
}\nonumber \\&&\times\lambda^{1/4}(1, \hat{m}_V^2,
\hat{m}_{K^*}^2)\, {A}_\lambda^V,
\end{eqnarray}
where ${\rm BR}(\ldots)$ is the branching ratio of $B_d^0 \to
K^{*0}\, V$ decay and $\tau_B$ is the lifetime of a $B$ meson.

Solving Eqs.~(\ref{eq:0A6}) we find the scalars $S_1\,,S_2$ and
$S_3$, and then extend the helicity amplitudes $A_\lambda^V$ off
the mass shell of the meson $V$, i.e. for $q^2 \ne m_V^2$. We
introduce the phases $ \delta_\lambda^V \equiv {\rm
arg}(h_\lambda^V)$, $ \delta_i^V \equiv {\rm arg}(S_i^V)$, where
$i=1\,,2\,,3$. Then we have
\begin{eqnarray} \label{eq:0A12}
|S_1^V|&=&\frac{|A^V_\||}{\sqrt{2}}\, ,\quad
 |S_3^V| = \sqrt{\frac{2}{\lambda(1,\hat{m}_V^2,\hat{m}_{K^*}^2)}}
\, |A_\perp^V|\,,
\nonumber \\
|S_2^V| & = &
\frac{\sqrt{2}}{\lambda(1,\hat{m}_V^2,\hat{m}_{K^*}^2)} \Bigl(
8\hat{m}_{K^*}^2 \hat{m}_V^2 |{A}_0^V|^2 \nonumber \\&&+
    \left(1-\hat{m}_V^2-\hat{m}_{K^*}^2\right)^2 |{A}_\parallel^V|^2
    \nonumber \\
   &&+
    4\sqrt{2}
    \hat{m}_{K^*}\hat{m}_V(1-\hat{m}_V^2-\hat{m}_{K^*}^2)\nonumber
    \\&&\times
    |{A}_0^V| |{A}_\parallel^V|\cos(\delta_\parallel^V - \delta_0^V)
    \Bigr)^{1/2} \,,
    \nonumber \\
\sin(\delta_2^V -\delta_0^V) &=& -
\frac{\sqrt{2}}{|S_2^V|\,\lambda(1,\hat{m}_V^2,\hat{m}_{K^*}^2)}\nonumber
\\&&\times
        (1-\hat{m}_V^2-\hat{m}_{K^*}^2)|{A}_\parallel^V|
        \sin(\delta_\parallel^V - \delta_0^V)\,,
     \nonumber \\
\cos(\delta_2^V -\delta_0^V) &=&  -
\frac{\sqrt{2}}{|S_2^V|\,\lambda(1,\hat{m}_V^2,\hat{m}_{K^*}^2)}\nonumber
\\&&\times
    \Bigl((1-\hat{m}_V^2-\hat{m}_{K^*}^2)|{A}_\parallel^V| \cos(\delta_\parallel^V -\delta_0^V)
    \nonumber \\&& +2\sqrt{2}\hat{m}_V \hat{m}_{K^*}
    |{A}_0^V|\Bigr)\,,\nonumber \\\delta_1^V&\equiv&\delta_\|^V \,(mod\, 2\pi) \, ,
\delta_3^V\equiv \delta_\perp^V \,(mod\, 2\pi) .
\end{eqnarray}


\end{document}